# Aspects of Polar-Coded Modulation

Mathis Seidl, Andreas Schenk, Clemens Stierstorfer, and Johannes B. Huber
Lehrstuhl für Informationsübertragung, Friedrich-Alexander-Universität Erlangen-Nürnberg, Erlangen, Germany,
email: {seidl, schenk, stierstorfer, huber}@lnt.de

*Abstract*—We consider the joint design of polar coding and higher-order modulation schemes for ever increased spectral efficiency. The close connection between the polar code construction and the multi-level coding approach is described in detail. Relations between different modulation schemes such as bit-interleaved coded modulation (BICM) and multi-level coding (MLC) in case of polar-coded modulation as well as the influence of the applied labeling rule and the selection of frozen channels are demonstrated.

## I. INTRODUCTION

The need for bandwidth-efficient coded modulation for modern communication systems is evident. This raises the question of the optimal combination of modulation and channel coding. To some extent this problem has been settled through the use of bit-interleaved coded modulation (BICM) [1]. In BICM coding and modulation are designed almost independently: a channel code optimized for the binary input AWGN channel is connected via a mapper and (possibly) an interleaver with a higher-order modulation scheme.

Compared to the optimal joint design of coding and modulation following the so-called multi-level coding (MLC) principle [2], BICM induces only a small loss in capacity. However, in particular for realistic finite-length codes, employing codes optimized for the binary-input AWGN channel instead of those optimized for the higher-order system at hand leads to significant performance degradation. This is well understood, e.g., for the application of convolutional codes, where channel codes optimized for trellis-coded modulation outperform conventional BICM approaches [3]. For other coding schemes, especially those mainly developed after Caire's fundamental paper on BICM in 1998 [1], such as turbo codes and low-density parity check codes, this problem has hardly been addressed.

The unique structure of polar codes [4] offers promising starting points for joint optimization of coding and modulation (so-called *polar-coded modulation*). This makes polar codes – if correctly designed – very attractive for coded modulation schemes, apart from their well-known benefits (in particular, the low-complexity encoding and decoding).

The aim of this paper is to show first results of and motivate a thorough approach to polar-coded modulation. To this end, we discuss and contrast the application of polar codes in MLC and BICM. We point out benefits and similarities from an information-theoretic conceptual point of view. Building on the relation of Gray and natural labeling, recently shown in [5], we proof the equivalence of polar-coded BICM and MLC for 4-QAM under the constraint of equal structural delay.

We underline the importance of a polar code construction optimized for the applied modulation scheme, i.e., in particular the optimized selection of so-called frozen channels. The observations are supported by means of numerical results. Based on these insights, we motivate points for further optimization of polar-coded modulation.

For sake of clarity and in order to facilitate our approach, we restrict ourselves to the additive white Gaussian noise (AWGN) channel. The transfer of insights gained for this setup to different channel models, especially incorporating fading scenarios, is an interesting task for future research.

This paper is organized as follows: After a brief review of the main building blocks in Sec. II, we discuss the application of polar codes in multi-level and bit-interleaved coded modulation in Sec. III and IV, respectively. Our main findings are summarized in Sec. V. The paper concludes with an outlook.

## II. BASIC BUILDING BLOCKS

First, we introduce the basic building blocks of polar-coded modulation, i.e., the higher-order pulse-amplitude modulation and channel coding using polar codes, focussing on the parts relevant for the application in coded modulation.

### A. Higher-Order Modulation

We consider the conventional discrete-time equivalent system model of $M$-ary digital pulse-amplitude modulation (PAM) [6] over the AWGN channel. The receive symbols are given as the transmitted amplitude coefficient[1] $x$ (taken from the signal constellation $\mathcal{X}$) corrupted by AWGN $z$ with variance $\sigma_Z^2$, i.e.,

$$y = x + z. \quad (1)$$

Each transmit symbol $x$ is addressed by a binary label $\boldsymbol{b}$ of $m = \log_2(M)$ bits, so-called *bit levels*. The mapping from binary labels to amplitude coefficients is specified by a bijective binary labeling rule $\mathcal{M} : \boldsymbol{b} \in \{0;1\}^m \mapsto x \in \mathcal{X}$ (e.g., Gray or set-partition (SP) labeling).

The overall rate of the coded modulation scheme is denoted as $R$. As a measure for the quality of the PAM channel, we consider the ratio of the transmitted energy per information bit $E_\text{b}$ over the one-sided noise power-spectral density $N_0$; we have $E_\text{b}/N_0 = E_\text{s}/(R \cdot N_0) = \sigma_X^2/(R \cdot \sigma_Z^2)$ (for carrier-modulated transmission, i.e., complex-valued $x$ and $z$), where $E_\text{s}$ denotes the energy per PAM symbol and $\sigma_X^2$ the variance of the transmit symbols.

---

[1]For compact notation, we omit explicit symbol/time indices, where it does not impair clarity.

## B. Polar Codes

Polar codes, recently introduced by E. Arıkan, have been shown to be the first channel coding construction that provably achieves the symmetric capacity of arbitrary binary-input discrete memoryless channels (B-DMCs) under low-complexity encoding and decoding [4].

*1) Code Construction:* Let W be a B-DMC and $I(\mathsf{W})$ its symmetric capacity, i.e., the mutual information of W assuming equiprobable binary source symbols. The encoding operation for a polar code of length $n$ may be described by multiplication of a length-$n$ vector $\boldsymbol{u}$ containing the information symbols with a generator matrix $\boldsymbol{G}_n$ that is defined by the recursive relation [4]

$$\boldsymbol{G}_n = \begin{bmatrix} \boldsymbol{G}_{n/2} & \boldsymbol{0} \\ \boldsymbol{G}_{n/2} & \boldsymbol{G}_{n/2} \end{bmatrix} \quad (2)$$

where $n \geq 2$ and $\boldsymbol{G}_1 = \begin{bmatrix} 1 \end{bmatrix}$. Clearly, by definition the block length is restricted to powers of two and encoding takes place over the binary field $\mathbb{F}_2$. The resulting codeword $\boldsymbol{c} = \boldsymbol{u}\boldsymbol{G}_n$ is then transmitted over the channel W (cf. Fig. 4).

The transmission of each source symbol $u_i$ can be described by its own binary-input channel $\mathsf{W}_n^{(i)}$. Throughout the paper, we will refer to these channels as *bit channels*, in analogy to the *bit levels* in coded modulation. The output of each channel $\mathsf{W}_n^{(i)}$ depends on the values of a specific set of symbols $u_0, \ldots u_{i-1}$, where the index $i$ corresponds to the coefficients of the information vector $\boldsymbol{u}$ in bit-reversed order. Thus, the channels $\mathsf{W}_n^{(i)}$ imply a specific decoding order.

The mutual information between the input $u_i$ and the output of the channel $\mathsf{W}_n^{(i)}$ is given by

$$I(\mathsf{W}_n^{(i)}) \stackrel{\text{def}}{=} I(U_i; \boldsymbol{Y}|U_0 U_1 \ldots U_{i-1}) . \quad (3)$$

The chain rule of mutual information assures

$$\begin{aligned} n \cdot I(\mathsf{W}) \stackrel{\text{def}}{=} I(\boldsymbol{C}; \boldsymbol{Y}) &= I(\boldsymbol{U}; \boldsymbol{Y}) \quad (4) \\ &= I(U_0; \boldsymbol{Y}) + I(U_1; \boldsymbol{Y}|U_0) + \ldots \\ &\quad + I(U_{n-1}; \boldsymbol{Y}|U_0 U_1 \ldots U_{n-2}) \\ &= \sum_{i=0}^{n-1} I(\mathsf{W}_n^{(i)}) . \end{aligned}$$

For data transmission only the bit channels with highest capacity are used, referred to as *information channels*. The data transmitted over the resting bit channels (so-called *frozen channels*) are fixed values which are known to the decoder. By this means, the code rate can be chosen in very small steps of $1/n$ without the need for changing the code construction.

With increasing block length, the set of bit channels $\mathsf{W}_n^{(i)}$ shows a polarization effect in the sense that the capacity $I(\mathsf{W}_n^{(i)})$ of almost each channel is either near 0 or near 1. The fraction of channels not being either completely noisy or completely noiseless tends to zero.

In order to select the optimal set of frozen channels, we need to compute/estimate the capacities $I(\mathsf{W}_n^{(i)})$. This can either be performed by simulation or by density evolution [7].

*2) Successive Decoding:* As the matrices $\boldsymbol{G}_n$ are self-inverse for all $\log_2(n) \in \mathbb{N}$, the estimation of $\boldsymbol{u}$ from $\boldsymbol{y}$ – the latter being a noisy version of the codeword $\boldsymbol{c}$ resulting from transmission over the channel W – is based on the very same scheme as for encoding. Of course, here information combining [9] of reliability values obtained from the channel output $\boldsymbol{y}$ is performed instead of $\mathbb{F}_2$ arithmetics. The successive cancellation (SC) decoding algorithm [4] for polar codes generates estimates on the information symbols $\hat{u}_i$ one after another (in bit-reversed order), making use of the already decoded symbols $\hat{u}_0, \ldots, \hat{u}_{i-1}$. As already noted by Arıkan [8] and elaborated in the following section, SC decoding of polar codes is in fact very similar to the multi-stage decoding process for multi-level coding.

## III. MULTI-LEVEL POLAR-CODED MODULATION

### A. Multi-Level Coding

The optimum combination of coding and modulation follows the MLC principle. Here, each bit level is encoded (hence, protected) individually using its own component code with correspondingly set code rate. The receiver then performs multi-stage decoding, i.e., computes reliability information for the first bit level, which is then decoded. This information is used for demapping and decoding of the next bit level, and so on.

The coded modulation [2], or constellation-constrained, capacity is given as the mutual information between the channel input and channel output assuming equiprobable source symbols, i.e.,

$$\begin{aligned} C_{\text{cm}} \stackrel{\text{def}}{=} I(X;Y) &= I(B_0 B_1 \ldots B_{m-1}; Y) \quad (5) \\ &= I(B_0; Y) + I(B_1; Y|B_0) + \ldots \\ &\quad + I(B_{m-1}; Y|B_0 B_1 \ldots B_{m-2}) \\ &= \sum_{i=0}^{m-1} I(\mathsf{B}_m^{(i)}) . \end{aligned}$$

We refer to the channels associated to the bit levels as $\mathsf{B}_m^{(i)}$, $i = 0, \ldots, m-1$. The second line follows from the chain rule of mutual information and can be interpreted as the parallel transmission of the binary label entries $b_i$, $i = 0, \ldots, m-1$, over $m$ memoryless binary input channels, followed by successive decoding [2], i.e., it resembles the multi-stage decoding process.

In order to increase the information transfer in multi-stage decoding between the bit levels, the bit-level capacity curves should be separated as far as possible [2]. This is usually achieved by using a binary labeling constructed according to the set-partitioning principle. Similar to polar codes, one aims for some kind of polarization within the bit levels in MLC.

### B. MLC Using Polar Codes

The MLC approach is closely related to polar codes on a conceptual level: First, note that multi-stage decoding is very similar to the successive decoding of polar codes from an information-theoretic point of view. Both decoders basically implement the well-known chain rule of the mutual information [10], cf. (4) and (5).

In MLC each bit level is decoded based on the (hard) decisions of the lower bit levels. If polar codes are employed as component codes, thus a first polar code is decoded, whose output is then fed to the demapper for the next bit level. Similarly, in the successive decoding process of polar codes, each bit is decided based on the feedback of all previous (hard) decisions.

Using MLC with multi-stage decoding, an $M$-ary channel is splitted into $m = \log_2(M)$ bit levels $\mathsf{B}_m^{(i)}$, $i = 0, \ldots, m-1$, that are B-DMCs if the underlying transmission channel is memoryless. Their symmmetric capacities sum up to $C_{\text{cm}}$, cf. (5). According to [4, Th. 1], the polar component codes approach each of these bit level capacities while their block length increases.

We thus conclude, that polar codes together with MLC and multi-stage decoding achieve the coded modulation capacity $C_{\text{cm}}$ for arbitrary $M$-ary signal constellations in case of a memoryless transmission channel. All results on the speed of convergence considering transmission over a single B-DMC hold as well in the case of MLC.

We remark that this (asymptotic) result is independent of the labeling strategy applied in MLC. However, for finite-length codes the labeling has significant impact on the performance of polar-coded MLC.

### C. Rate Allocation

At this point we note a further benefit of applying polar codes in MLC: the MLC approach requires to flexibly select codes of various rates for each bit level. For ideal component codes, the optimum rate allocation follows the so-called capacity rule [2], i.e., the code rate of the component code for the $i$th bit level should be set according to its bit-level capacity $I(\mathsf{B}_m^{(i)})$, cf. (5).

In multi-level polar-coded modulation, however, a predefined rate allocation is actually not required. It is sufficient to compute the bit-channel capacities $I(\mathsf{W}_n^{(i)})$ for the $m$ polar codes used in the multi-level coded system under investigation at a given $E_s/N_0$ operating point (e.g., simply via simulation). Since in MLC we apply multi-stage decoding, these $m$ polar codes are decoded in a successive manner, making use of the decisions of the previous results. Hence, we may virtually concatenate the bit channels of the $m$ polar codes in a single bit-channel-capacity plot in a similar fashion as for the well-known polarization plots [4]. Selecting the frozen channels out of all bit channels according to the desired overall code rate of the MLC scheme implicitly allocates a different number of frozen channels for each bit level.

Due to the equivalence of successive-cancellation and multi-stage decoding, this allocation basically implements the capacity rule. However, here, the $E_s/N_0$ operating point and the finite codeword length are inherently taken into account, as well.

Exemplarily considering 4-ASK at $10 \log_{10}(E_s/N_0) = 2.1$ dB (yielding $C_{\text{cm}} = 1$), the resulting bit-channel capacities are shown in the top part of Fig. 1 (length-128 component codes). Designing for $R = 1$ (corresponding to the $E_s/N_0$ operating point of $2.1$ dB) we obtain a rate allocation of $24/128$ and $104/128$ for the two bit levels, respectively. For this operating point, the capacity rule gives $0.1951$ and $0.8049$, respectively. After quantization of these rates to the available rates of a length-128 polar code, both rate allocations coincide.

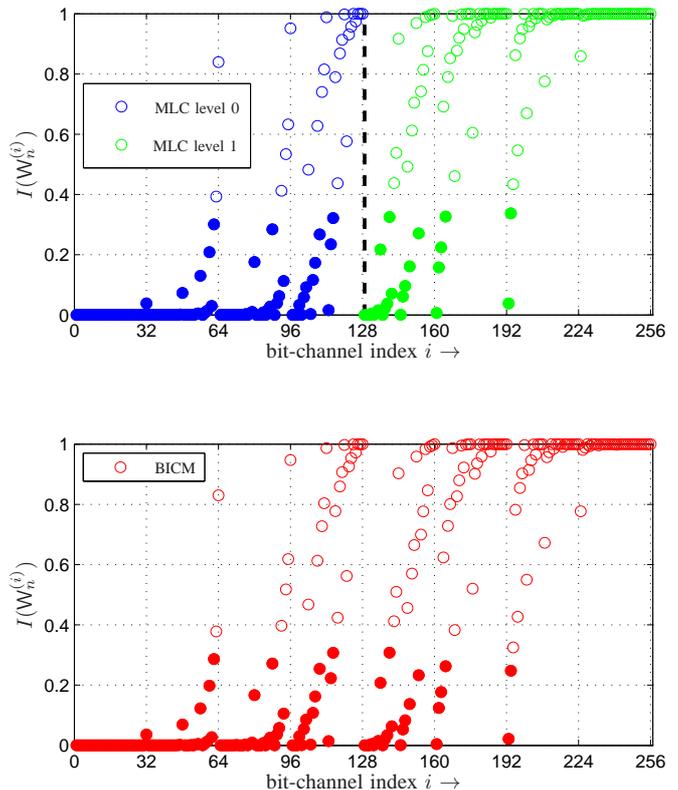

Fig. 1. Bit-channel capacities of polar-coded modulation using 4-ASK. Frozen channels for a design at $R = 1$ are demarked by filled circles. Top part: MLC using SP-labeled ASK with component codes of length 128. Bottom part: BICM using Gray-labeled ASK with code of length 256. $10 \log_{10}(E_s/N_0) = 2.1$ dB.

## IV. BIT-INTERLEAVED POLAR-CODED MODULATION

### A. BICM

As opposed to MLC, in a BICM setup all bit levels are treated equally at both sides, the transmitter and the receiver [1], [11]. The source bits are simply encoded using a single rate-$R_c$ binary channel code. The code symbols are (possibly) interleaved according to some pseudo-random order and partitioned into $m$-touples $\boldsymbol{b}$ of code symbols, which are mapped to amplitude coefficients $x$ (thus, $R = R_c m$).

The BICM receiver performs parallel decoding, i.e., it neglects the relations between the bit levels and computes reliability information independently for each bit level based on the received symbol. These bit metrics are deinterleaved and fed to the decoder.

Similar to polar-coded MLC, in the SC decoding of the polar code each bit is decided based on all previous decisions.

However, as opposed to polar-coded MLC, this information is used in the decoder only and not already in the demapper. This causes the loss of BICM vs. MLC.

The BICM capacity is given as the sum of the bit level capacities $I(B_i;Y)$ neglecting the feedback of lower bit levels; therefore, it is usually smaller than the coded-modulation capacity:

$$C_{\text{bicm}} \stackrel{\text{def}}{=} \sum_{i=0}^{m-1} I(B_i;Y) \leq C_{\text{cm}}. \qquad (6)$$

Since each bit level is treated equally, as opposed to the MLC approach, the capacities of the bit levels should be close to each other, i.e., we aim for a low amount of polarization within the bit levels. This is achieved using Gray labeling.

### B. Selection of Frozen Channels

One open problem of polar-coded BICM is to determine the optimal set of frozen channels in the code construction. To this end, we distinguish two strategies:

The straight-forward and probably most common approach is to simply use a selection optimized for binary ASK transmission, i.e., the same set of frozen channels as obtained when computing the bit-channel capacities of a binary-input AWGN channel. This option is valid for the case of a pseudo-random interleaver in BICM, as effectively this interleaver removes the memory of the constellation-inherent fading process, which allows to assume that each coded bit on average sees the same channel. Since this memory is neglected, on an AWGN channel this strategy leads to non-optimal performance. Moreover, it is an open problem, which equivalent binary ASK channel should be chosen to model the considered higher-order modulation, i.e., in particular which $E_s/N_0$ should be used.

Instead, for BICM over the AWGN channel, no interleaver should be applied between channel code and mapper. Thus, the set of frozen channels may directly be obtained from computing the bit-channel capacities for the applied modulation scheme. In doing so, the polar code optimally adapts to the constellation-inherent fading of the modulation scheme under investigation, i.e., the frozen channels are allocated to the different bit levels in an optimum order[2]. Clearly, this selection of frozen channels depends on the operating point, i.e., the ratio $E_s/N_0$, too.

Exemplarily, for the same setup as in MLC, the resulting set of frozen channels is shown in the bottom part of Fig. 1, obtained from training for 4-ASK and $n = 256$. We observe a very similar shape compared to the MLC bit-channel capacities. However, each value is slightly lower for BICM due to the loss in capacity of BICM vs. MLC, cf. (6).

[2]This process can be viewed as a permutation of code symbols prior to mapping, i.e., the application of a well-determined non-random interleaver in the spirit of [12]. Moreover, please note that if the values transmitted over the frozen channels are set to zero, for some specific settings this results in an increased transmit power, which of course has to be taken into account in the definition of $E_s/N_0$.

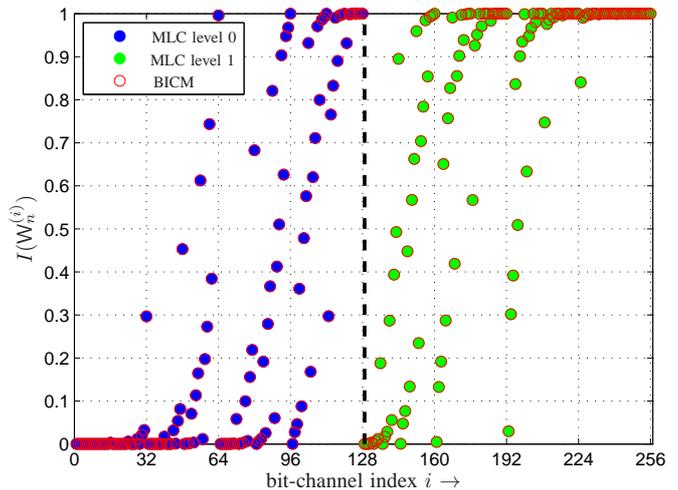

Fig. 2. Bit-channel capacities for polar-coded BICM with Gray-labeled 4-QAM (codeword length 256) and an polar-coded MLC with SP-labeled 4-QAM (length-128 component codes) at $10 \log_{10}(E_s/N_0) = 1$ dB.

### V. COMPARISON

In order to enable a fair comparison of MLC and BICM, we fix the structural delay [13] of both coded modulation schemes. Thus, for a codeword length in BICM of $n$, each component code in MLC must have a codeword length of only $n/m$.

### A. The Special Case 4-QAM

Clearly, in 4-QAM there is no difference in capacity of a Gray-labeled BICM approach and MLC, due to the orthogonality of the inphase and quadrature component and the separability of the labeling rule. However, for finite codeword lengths the MLC approach potentially suffers a performance degradation in bit error rate. This is due to the constraint of equal structural delay, i.e., for a codeword length of $n$ in BICM, the codeword length of the component codes in MLC is restricted to $n/2$. The multi-stage decoding in MLC thus has to compensate for the reduced codeword length compared to parallel decoding in BICM. A general statement on the exact relative performance of MLC and BICM using similar codes under this constraint is hardly possible.

For the case of polar-coded 4-QAM, however, both approaches perform exactly equal, i.e., all bit-channel capacities and the bit error rate performance are exactly equal, cf. Fig. 2, which emphasizes the strong relation of the successive decoding of polar codes and multi-stage decoding in MLC.

This can be explained making use of the connection of Gray and set-partitioning labeling, as recently pointed out in [5]. Collecting all possible binary labels as the columns of a matrix, we have

$$\boldsymbol{M}_{\text{SP}} = \begin{bmatrix} 0 & 0 & 1 & 1 \\ 0 & 1 & 0 & 1 \end{bmatrix} \text{ and } \boldsymbol{M}_{\text{Gray}} = \begin{bmatrix} 0 & 0 & 1 & 1 \\ 0 & 1 & 1 & 0 \end{bmatrix}$$

for the SP and Gray labeling, respectively. Both can be transformed into each other using the matrix $\boldsymbol{T}$, i.e., $\boldsymbol{M}_{\text{SP}} =$

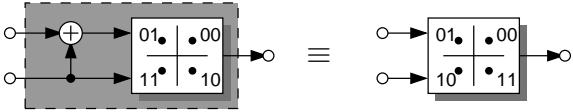

Fig. 3. Transformation of Gray labeling into SP labeling for 4-QAM.

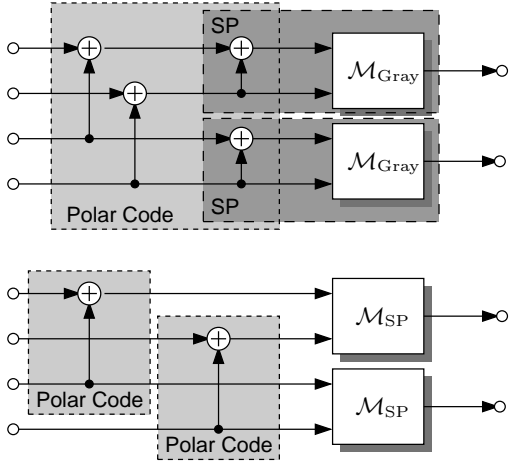

Fig. 4. Block diagram of polar-coded BICM (top part) for length-4 code and Gray labeling and MLC (bottom part) for length-2 component codes and SP labeling. Highlighted: Equivalence of first stage in BICM to SP labeling.

$T^{-1}M_{\text{Gray}}$ and $M_{\text{Gray}} = TM_{\text{SP}}$, where

$$T = \begin{bmatrix} 1 & 0 \\ 1 & 1 \end{bmatrix} \quad (7)$$

with the operations conducted over the Galois field $\mathbb{F}_2$. This processing is shown in Fig. 3.

Making use of this relation, it can be shown that a polar-coded Gray-labeled BICM system with codeword length $n$ is equivalent to a polar-coded SP-labeled MLC system with component codes of length $n/2$. This is depicted for the toy-example $n = 4$ in Fig. 4. The first stage of the polar code in Gray-labeled BICM is equivalent to SP labeling. Hence, each bit level is protected by its own polar code of length 2, as in the MLC setup.

Due to the recursive definition of polar codes this result directly extends to $n > 4$ (with $\log_2(n) \in \mathbb{N}$), but only holds for $m = 2$ and equal BICM and MLC capacity, i.e., 4-QAM. For this special case the transformation matrix is self-inverse and equals the matrix underlying the construction of polar codes, cf. (2). Unfortunately, for $m > 2$ this property is no longer valid. The matrix $T$ differs from that of the polar code (and is, in general, no longer self-inverse). However, it is an interesting point for future research to adopt the labeling rule or equivalently adopt the polar code generator matrix.

### B. Selection of Frozen Channels in BICM

Finally, we address the problem of selecting the frozen channels for code construction of BICM using $M$-ary ASK.

As described in Sec. IV-B, there are basically two different options. On the one hand, we can use an allocation optimized for the binary AWGN channel in combination with a pseudo-random interleaver. On the other hand, we can directly design the polar code for the considered ASK system, in which no interleaver is applied. Note that for both variants a desired operating point of $E_{\text{s}}/N_0$ has to be specified.

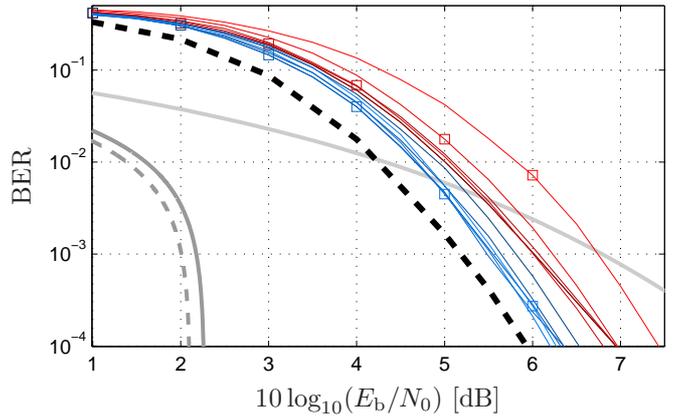

Fig. 5. BER of polar-coded BICM using Gray-labeled 4-ASK at $R = 1$ bit/symbol (length-256 polar code) for different selections of the frozen channels (blue: optimized for 4-ASK transmission at indicated operating point, red: optimized for binary AWGN at indicated operating point) compared to polar-coded MLC (dashed-black, SP labeling), rate-distortion bound (solid-gray: BICM, dashed-gray: MLC) and uncoded BPSK (light gray).

Fig. 5 depicts the bit error rates of polar-coded BICM using Gray-labeled 4-ASK and $R = 1$ (length-256 codes) for the two variants optimized to different $E_{\text{s}}/N_0$ operating points (from 1 to 6 dB, indicated by a marker). The performance of polar-coded MLC (SP-labeling, length-128 component codes), uncoded binary ASK (with equal rate), and the rate-distortion bounds of BICM and MLC are shown for reference.

The set of frozen channels optimized to the system at hand significantly outperforms the setup, where the polar code has been designed for a binary AWGN channel. Additionally, the latter option is very susceptible to the selection of frozen channels. In particular, at a given operating point best performance is not achieved when the system is trained for this point. For the first option, however, the best performance is achieved when the system is trained – i.e., the set of frozen channels is optimally chosen – exactly for the desired operating point, as expected.

## VI. Summary and Outlook

In this paper we have given a structured approach to polar-coded modulation. We have discussed similarities and equivalences on a conceptual level. These insights have been utilized to optimize polar-coded BICM and MLC.

Several points remain open for optimization of polar-coded modulation. A starting point for this optimization is the selection of frozen channels in the polar code construction matched to the higher-order constellation, as motivated in this paper. To this end, we believe that the similarity of the transformation of Gray into SP labeling with the polar code construction can be exploited beneficially. Finally, the incorporation of fading channels appears to be an important task for future work.